\documentclass{aa}
\usepackage{graphicx}
\begin{document}
\title{Solar active regions: a nonparametric statistical analysis}
   \author{J. Pelt
          \inst{1}
          \and
          M. J. Korpi
          \inst{2}
          \and
          I. Tuominen
          \inst{2} 
          }
   \offprints{J. Pelt}
   \authorrunning{J. Pelt et al.}
   \titlerunning{Solar active regions} 
   \institute{
             Tartu Observatory, 61602 T\~{o}ravere, Estonia
             \and
             Observatory, PO Box 14, FI-00014 University of Helsinki, Finland           
             }
             
             \date{Received ---; accepted ---} 
       
\abstract{The sunspots and other solar activity indicators tend to cluster on the surface of the Sun. These clusters very often occur
   at certain longitudes that persist in time.
   It is of general interest to find new and simple ways to characterize the observed distributions of different indicators and their behaviour in time.}
  {In the present work we use Greenwich sunspot data to evaluate statistical but not totally coherent stability of sunspot distribution along latitudes as well as longitudes. 
The aim was to obtain information on the longitudinal distribution of the underlying spot-generating mechanism rather than on the distribution 
and migration of sunspots or sunspot groups on the solar surface. Therefore only sunspot groups were included in the analysis, and only the time of 
their first appearance was used. }
  {We use simple nonparametric approach to reveal sunspot migration patterns and their persistency.} 
  {Our analysis shows that regions where spots are generated tend to rotate differentially as the spots   and spot groups themselves do. 
   The activity areas, however, tend to break down relatively fast, during 7-15 solar rotations.}
  {This study provides a challenge for solar dynamo models, as our results are consistent with the presence of a non-axisymmetric spot-generating mechanism experiencing differential rotation (known as phase mixing in dynamo theory). The new nonparametric method introduced here, completely independent of the choice of the longitudinal distribution of sunspots, was found to be a very powerful tool for spatio-temporal analysis of surface features.}

\maketitle
\section{Introduction}\label{Section:introduction}
Modern observations of the Sun are so rich in detail that astronomers
are eventually stroken by ``embarrassment of riches''.  When
spatio-temporal properties of the smaller features - say spots,
flares {\it etc} - are treated with well-established vigour, the
analysis of spatially larger or temporarily longer patterns is 
very complicated. Even the nomenclature of the phenomena is not
well-established - for instance the time-space cluster of the
local phenomena can be called as ``active
longitude''(Losh~\cite{Losh}, Vitinskij~\cite{Vitinskij}),{\it
  ``Sonnefleckenherd''} (Becker~\cite{Becker}), ``active region''
(Bumba \& Howard~\cite{Bumba65}), ``sunspot nest'' (Castenmiller et
al.~\cite{Castenmiller}) , ``complex of activity'' (Gaizauskas et
al.~\cite{Gaizauskas83}) or ``hot spot'' (Bai~\cite{Bai88}). 
There is, in addition, a problem with proper definition of such
extended patterns.

It is generally thought that the tracers of solar activity - sunpots,
flares {\it etc} - are randomly generated manifestations of 
the larger scale mean magnetic field of the Sun generated by a
hydromagnetic dynamo process.  
An analogy with a submerged animal blowing out bubbles is quite
appropriate in this context (see Bai~\cite{Bai03}). What can we tell
about the swimming speed and size of the animal, if only random
bubbles are observable? How deep in water is the animal?

The answers to these kind of questions depend very much on the method
of analysis used. Very often subjective judgement is involved,
either through steps of visual processing or through involvement of
freely chosen procedure parameters (bin sizes, zone widths, detection
limits, preselection criteria {\it etc}).
 
From the statistical analysis point of view we can divide the
previously used methods along two lines: how the input data is
transformed before computing final statistics and what kind of
statistics are used. Some typical but random examples:
\begin{itemize}
\item Aggregated data (daily Wolf numbers) and correlation analysis
  (Bogart~\cite{Bogart}),
\item Raw heliographic longitudes and longitude-wise binning (Trotter \& Billings~\cite{Trotter},
Warwick~\cite{Warwick}),
\item Transformed (using trial rotation velocity) longitudes and $\chi
  ^2$ statistic (Bai~\cite{Bai87}),
\item Transformed (using latitude-dependent rotation velocities)
  longitudes and pattern matching (Usoskin et al.~\cite{UBP}; Pelt et
  al~\cite{PBKT}, hereafter PBKT),
\item Spherical harmonic decomposition and time series analysis of
  mode amplitudes, phases and phase-walks (Juckett~\cite{Juckett03}).
\end{itemize}
The third important aspect of the analysis is time coverage of the
observations. It is quite easy to find recurrent patterns in short
time series, but coherence tends to break down very fast for longer
datasets.

In this paper we in a certain sense try to return to the square one,
back to the very basics. 
Using very simple considerations and
avoiding all freely chosen parameters we try to get answers to the
questions:
\begin{itemize}
\item Is there a tendency for surface elements to occur at
  certain longitudes that persist over time?
\item How this persistence and differential rotation of the surface
  elements are connected?
\item How long typical correlations in activity persist?
\end{itemize} 
Our aim is not so much to perform another statistical analysis of the
well-known and already extensively analysed data, but to introduce a
new nonparametric method of analysis involving no
physical, geometrical or statistical prior assumptions. 
In Section~\ref{method} we introduce our method of analysis, in
Section~\ref{analysis} we present the results obtained for the
Greenwich sunspot data set, and finally in Sections~\ref{discussion}
and \ref{conclusions} we discuss our results in the light of previous
statistical analyses.

\section{Method of analysis}\label{method}

\subsection{Nonparametric method}
Let us assume that we have two sets of longitudes:
$\lambda_{i}^{(1)},i=1,\dots,N$ and
$\lambda_{j}^{(2)},j=1,\dots,M$. 
Their values belong to the interval
$0^{\circ}\le \lambda \le 360^{\circ}$ and we assume that $N<=M$ (if otherwise, we can
always swap the sets). 
We want to characterize the similarity or
the difference between the two longitude distributions somehow. 
The general theory
of directional measurements is considered in mathematical statistics
(see for instance the latest monograph by Mardia \& Jupp~\cite{Mardia} and
references therein), but here we need a more specific method, namely one without
any underlying statistical assumptions or parametric models for
the distributions involved.

We propose the following very simple nonparametric method:
The circular distance between two longitudes
$\lambda_k$ and $\lambda_l$ we define as usually done
\begin{equation}
\Delta \lambda _{k,l}  = \min (\left| {\lambda _k  - \lambda _l } \right|,360^ \circ   - \left| {\lambda _k  - \lambda _l } \right|).
\end{equation}
Let us take a particular longitude $\lambda_{k}^{(1)}$ from the first
set. Among the longitudes of the second set there is always a value
whose circular distance from the selected value is the smallest, let us
denote this distance as $\Delta \lambda_k$. All together we can
compute $N$ such values - for each longitude in the first set. Now we
select the longest distance among them and denote it simply as
$\Delta$. It is quite clear that, in the particular case when the first set
is just a subset of the second one, $\Delta= 0$. If the sets differ, then
$\Delta>0$. In principle such max-min distance between two longitude
sets is already a useful statistics; its full power, however, is revealed if we
properly normalize it.

For a particular set sizes $N$ and $M$ we can compute the mathematical
expectation of $\Delta$ for completely random distributions of
longitudes in both sets. Let us denote this expectation as $\bar
\Delta$. Our final statistics which measures statistical distance
between the two sets of longitudes is then:
\begin{equation}
D = \frac{\Delta }{{\bar \Delta }}.
\end{equation}
If we want to stress that the distance $D$ is computed for two particular
indexed longitude sets, say for index $n$ ($N$ longitudes) and index
$m$ ($M$ longitudes) then we use notation
\begin{equation}
D(n,m) = \frac{{\Delta _{n,m} }}{{\bar \Delta _{N,M} }}.
\end{equation}
The mathematical expectations ${\bar \Delta _{N,M}}$ depend only on
the integers $N$ and $M$, and can be pretabulated.  In our calculations we
used 
approximations obtained from randomly generated longitudes for $10000$ statistically
independent runs.

It is quite obvious that for absolutely random pairs of longitude sets
our distance will have a value around $1$. For weakly correlated sets
values are less than one, 
and values higher than $1$ can occur when the longitude sets involved
are constrained in a certain way due to which they cannot form all
the patterns which occur for randomly generated sets. In the
case of sunspot groups, for instance, the distributions are constrained
by group sizes. Randomly generated points can fall arbitrarily
close, which is not true for sunspot longitudes, because for them the
group centres are separated by definition.

Having now the statistic to measure distances between different
distributions of longitudes we can go further. For a sequence of
longitude sets we can compute a mean distance between neighbouring
sets:
\begin{equation}
\bar D = \frac{{\sum\limits_{k = 1}^{K - 1} {D(k,k + 1)} }}{{K - 1}},
\end{equation}
where $K$ is the number of the sets. 
We can also investigate how the distance depends on
the mutual positions of particular sets:
\begin{equation}
\bar C(l) = \frac{{\sum\limits_{k = 1}^{K - l} {D(k,k + l)} }}{{K - l}}.
\end{equation}
Eventually $\bar D = \bar C(1)$. The statistic $\bar D$ allows us to
investigate rotational properties of the sunspot groups and statistic
$\bar C(l)$ will be used to estimate how persistent the longitudinal
correlations are.

\subsection{Rotation and frames}

Heliographic longitudes are defined using the so called Carrington frame,
which rotates against fixed stars with the exact period of $P_{C}=25.38$
days. The mean rotation period if observed from the Earth is
$P_{O}=27.2753$ days. The Carrington frame is a formal construct and
real features on the Sun need not to follow it exactly.

Let us fix 
a certain longitude $\lambda^{(C)}$ of a
particular persistent feature on the Sun rotating with the Carrington
angular velocity. Then its longitude for different Carrington
rotations $i$ will be fixed: $\lambda_i^{(C)} = \left[ {\lambda^{(C)}
    + i \times 360^{\circ}} \right] = \lambda^{(C)}$, angular brackets 
denoting here and below reduction to the interval $(0,360^{\circ})$. Because the
angular velocity of the Carrington frame is $\Omega_C =
\frac{{360^{\circ}}}{{P_C }}$ degrees per day we can rewrite cycle dependent
sequence of longitudes as
\begin{equation}
\lambda_i^{(C)}  = \left[ {\lambda^{(C)}  + i \times \Omega _C P_C } \right],i = 0,1, \ldots. 
\end{equation}
The actual angular velocity of an arbitrary feature on the Sun need
not to be exactly $\Omega_C$.  Let longitude of the first occurrence
of such feature be $\lambda$. Then cyclic reoccurrences of it can be
described using a correcting term $\Delta \Omega$:
\begin{equation}
\lambda _i  = \left[ {\lambda  + i \times (\Omega _C  + \Delta \Omega )P_C } \right],i = 0,1, \ldots. 
\end{equation}
The corrected frame rotates against Carrington frame with angular
velocity $\Delta \Omega$ degrees per day.  For convenience we
introduce also notion for siderial angular velocity of the accelerated
or decelerated frames $\Omega=\Omega_{C}+\Delta \Omega$.

In what follows, we measure the angular velocity in degrees per day,
latitude in degrees, and periods in days, and give the values in these
units.

\subsection{Algorithms}

All ingredients of the method of analysis described, we can now
formulate our basic algorithms.

As an input data we use a set of time tagged longitudes
$t_l,\lambda_l,l=1,\dots,L$, amounting to $L$ pairs of data.
Using the time points $t_l$ we divide records
into subintervals with the length $27.2753$ (Carrington rotations). This
procedure is not absolutely exact because the observation timing
depends on the somewhat excentric orbit of the Earth. Fortunately the
errors involved are small and we can ignore them. From the point of
generality and objectivity our choice is quite natural. Historical
observations are all done from Earth and consequently the features can
be observed only half a time. However, during the Carrington rotation
we can record what happens at all longitudes. As far as timing is
considered, due to the rotation some processes can actually start
earlier than observed. This excludes short-living processes (shorter
than Carrington rotation) from our analysis.

It is also possible to divide observations into longer
subintervals. Then we increase statistical stability of our estimates
(more observations in subsets) but loose resolution in time. We
consider time step with the length of one Carrington rotation to be
optimal.

We assume that the features on the surface of the Sun rotate with angular
velocity which is different from the Carrington velocity $\Omega_{C}$. For
a certain trial angular velocity $\Omega$ and for each Carrington
cycle $i$ we can compute longitude corrections:
\begin{equation}
\Lambda_i  = i \times \Delta \Omega P_C  = i \times (\Omega  - \Omega _C )P_C ,i = 0,1, \ldots. 
\end{equation}
By substracting rotation number dependent corrections from measured
longitudes and properly reducing results to interval $(0,360)$ we
build transformed longitudes:
\begin{equation}
\lambda^{(T)}=\left [ \lambda_i-\Lambda_i \right ].
\end{equation}
They can be analysed using the statistics introduced
above. We can also say that we transform longitudes in the Carrington
frame into longitudes in the comoving frame. The frame rotation velocity
$\Omega$ is a free parameter of the procedure. We expect that
the distributions of the transformed longitudes depend on $\Omega$ and the
highest level of correlation in the longitude distribution will show up as a 
minimum of the distance statistic $\bar{D}$.

First we compute how the mean distance between neighbouring rotations
$\bar D$ depends on angular velocity $\Omega$. Then we can use the
best value (producing the highest level of correlation) for angular
velocity to compute how distances depend on the interval between rotations
(using the statistic $\bar C(l)$).
     
\section{Data analysis}\label{analysis}
Here we describe how we apply the presented statistical method to
study the particular case of sunspots.

The most comprehensive (in time) compilation of sunspot data was
downloaded from the Science at NASA web
site\footnote{http://solarscience.msfc.nasa.gov/greenwch.shtml}.
The same minor corrections as in PBKT
were introduced.  In this paper we used the full data set
covering years 1874-2008, or in terms of Carrington rotations, the
rotations 275-2074. From all the data base records we chose only
sunspot groups, leaving out single spots. In this way all the entries
in the final set have equal statistical weight. For each sunspot group
we selected only the record of its first occurence. This is important
aspect of our analysis. We do not track sunspots as they rotate, but
are interested in the movement of the underlying spot-generating structures.
The final compiled data sets cover rotations 275-2074 with 16053
records for the Northern hemisphere and 275-2071 rotations with 15858
records for the Southern hemisphere; the compiled data sets are
available on the web\footnote{http://www.aai.ee/\~{}pelt/soft.htm}.

\subsection{Mean angular velocity}
\begin{figure}
  \resizebox{\hsize}{!}{\includegraphics{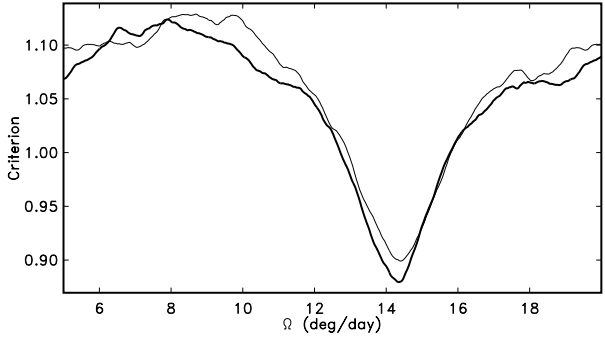}}
  \caption{Statistic $\bar D(\Omega)$ for the full dataset of the
    Northern hemisphere (thick line) and for the full dataset of the
    Southern hemisphere (thin line). Minima indicate the best fitting
    comoving velocities.}
  \label{Fig001}
\end{figure}

For the first approximation we can assume that the mechanism
generating the sunspots rotates as a rigid body. 
Then we can measure its angular velocity using $\bar D$ statistic
by comparing different longitude correction schemes,
and choosing the one that produces the lowest value of the statistic.
The results can be
best illustrated by displaying $\bar D$ as a function of $\Omega$ -
the actual siderial angular velocity of the frame. In Fig.~\ref{Fig001}
such functions are displayed for both solar hemispheres. As we see
both curves show very clear and indicative minima. The absolute
mininimum for the Northern data is positioned at 14.348 and for
Southern data at 14.403. The curves themselves are somewhat
fluctuating so that we found useful to estimate the minima using local
fits of the fifth-degree polynomials also. The resulting values from
the fitting procedure are $14.365\pm0.001$ and $14.409\pm0.002$ for
North and South, deviating from the absolute minimum values by 0.017
and 0.006, respectively.

From these results we can see that the Southern part of the mechanism
tends to rotate slightly faster. However the difference is too small,
especially if to take into account the roughness of our method, to be
conclusive.

\subsection{Differential rotation} \label{drot}

\begin{figure}
  \resizebox{\hsize}{!}{\includegraphics{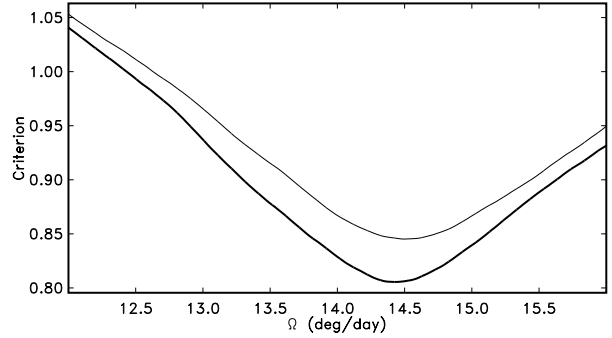}}
  \caption{Statistic $\bar D(\Omega)$ for the latitude strip
    $0.0-9.8$. Thick line - Northern hemisphere, thin line Southern
    hemisphere. Minima indicate the best fitting comoving velocities.}
  \label{Fig002}
\end{figure}

Sunspots and other activity indicators rotate with different angular
velocities at different latitudes. 
By tracking particular objects in
time it is possible to build a smooth curve to reveal the overall
pattern of such differential, latitude dependent, rotation. Our
statistic $\bar D(\Omega)$ does not track single sunspots or the
actual movement of sunspot groups, as we include only the first
appearance of the sunspot groups. 
This way we can check whether the spot-generating mechanism itself
rotates differentially or not. For that purpose we divided the observed
groups into four subsets along latitudes (per hemisphere) and computed
$\bar D(\Omega)$ for every group. The latitude limits for the subsets
where chosen to make them as equal in size as possible. The typical
curves are shown in Fig.~\ref{Fig002}.  The exact determination of the
minima for the curves is somewhat complicated. If we locally fit
polynomials into the curves as we did above, we can get estimates with
high formal precision (0.001-0.002).
The differences between the absolute numerical minima and the
fitted minima, however, can be quite large (up to 0.045). 
Therefore we can claim that the probable statistical errors of the
obtained minima are around 0.02 degrees per day.

\begin{table}
\centering
\begin{tabular}{lcccc} 
\hline
\hline
Hemisphere & Latitude & N &Angular &$T_{\rm bd}$  \\
&range & &velocity &\\
\hline
North & 0.0-9.8 & 4013  & 14.430 &80.1\\
      & 9.9-14.4 & 4015  & 14.224 &71.8\\
      & 14.5-20.0 & 4048  & 14.371 &42.4\\
      & 20.1-60.0 & 3977  & 13.974 &3.7\\
South & 0.0-9.8 & 4095 &  14.496 &78.3\\
      & 9.9-14.4 & 3847 &  14.423 &74.8\\
      & 14.5-20.0 & 4143 &   14.370 &41.3\\
      & 20.1-60.0 & 3809 &  13.964 &3.9\\
\hline
\end{tabular}
\label{Tab1}
\caption{Differential rotation}
\end{table}

The full set of the absolute minima for all the eight curves is given in
Table~\ref{Tab1}. 
\begin{figure}
  \resizebox{\hsize}{!}{\includegraphics{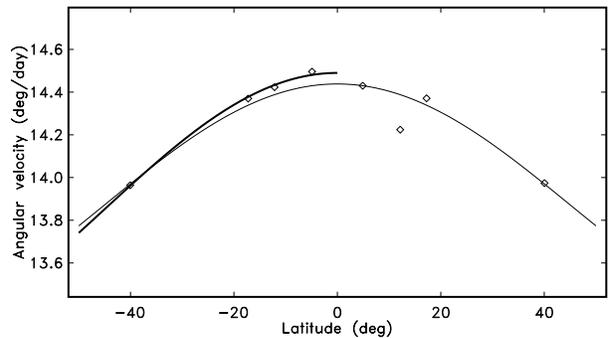}}
  \caption{Differential rotation curve. Thick curve - fit for Southern data, thin curve - full fit.}
  \label{Fig002b}
\end{figure}
In Fig.~\ref{Fig002b} the traditional least squares fit of the obtained
angular velocities $\Omega_i$ is presented. We estimated parameters $A$ and $B$ for a simple model:
\begin{equation}
\Omega_i  = A + B\sin ^2 (\theta_i ),
\label{Eq1}
\end{equation}
where $\theta_i$-s are mid-latitudes of the eight belts. The estimated parameters
were $A=14.438\pm 0.035$ and $B=-1.13\pm 0.17$ correspondingly.
For the Southern hemisphere the latitude dependence of the angular
velocity is monotonically decreasing polewards and resembles quite
well the curves obtained from sunspot tracking. For the Northern
hemisphere the latitudinal behaviour is not monotonous, and the two
latitude bands nearest to the equator are somewhat slower than
expected. 
If we used only Southern velocities for the fitting procedure, the resulting
fit was better with parameters $A=14.490\pm 0.001$ and $B=-1.276\pm 0.044$. At the moment we do not try to find any statistical or
physical interpretation of this result; more importantly to us, the
results obtained in this section clearly demonstrate that the simple
nonparametric method can be succesfully used to study the differential
rotation of the solar activity tracers.

\subsection{Break down times}
The results of the previous section clearly show that the
longitudinally concentrated spot-generating mechanism is subject to
differential rotation. Kinematic mean-field dynamo theory predicts
(e.g. Krause\&R\"adler 1980) that in the parameter regime where
nonaxisymmetric dynamo modes can be excited, the nonaxisymmetric modes
are non-oscillatory and rotate rigidly with angular velocity different
from the overall rotation period. The phenomenon of phase-mixing,
i.e. the nonaxisymmetric modes becoming affected by differential
rotation, is against these predictions; our results, however, are
consistent with the phase-mixing effect. 

Let us now try to quantify the effect of the differential rotation on
the nonaxisymmetric structures by calculating the characteristic time
needed to break down a longitudinally elongated stucture for different
latitude strips.
Using the estimated $B$ values from Eq.~\ref{Eq1} we can define a break-down time for the strip 
of latitudes $(\theta_1,\theta_2)$:
\begin{equation}
T_{\rm bd}  = \frac{{\Delta \phi }}{{B(\sin ^2 \theta _1  - \sin ^2 \theta _2 )}},
\end{equation}
where $\Delta \phi$ is the phase distance over which the hypothetical
longitudinal pattern can be regarded to be destroyed.  A reasonable
value for the parameter $\Delta \phi$ comes from a following
simple observation. Let us assume that at the latitudes $\theta_1$ and
$\theta_2$ we have $K$ observations. Let the longitudes coincide for
the starting point in time. The statistics introduced above are based on
finding the nearest ``neighbours''. In our case for every observation at
one latidude there is exactly one ``neighbour'' at the other. To break
the ties between the neighbours we need to rotate the observation at the other
latitude. That means we need to have relative shifts which are longer
than half the distance between two consequtive observations or
formally $\Delta \phi=0.5 \times {360}^{\circ}/\bar K$, where $\bar K$
is a mean number of observations per Carrington rotation.  Applying
all this to the same latitude intervals as in Sect.~\ref{drot}, we
obtain the break up times listed in Table~\ref{Tab1}.
   
\subsection{Decorrelation time}

So far we have demonstrated that the sunspot group distributions along
longitudes for sequential Carrington rotations are correlated. 
We also computed the best fitting mean angular velocity $\Omega$ for
several latitude strips, and as a result found out a clear
differential rotation pattern.  
Next we are interested in estimating the approximate lifetimes of the
correlated features found from the sunspot data.
For that purpose we use the obtained mean angular velocities for
each latitude strip and compute $\bar C(l)$ curves to learn how fast
the correlation between appropriately rotated longitude sets fades off.
\begin{figure}
  \resizebox{\hsize}{!}{\includegraphics{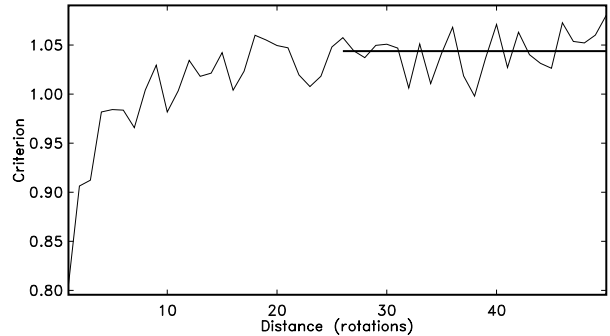}}
  \caption{Statistic $\bar C(l)$ computed for the Northern latitude strip
    $0.0-9.8$. Horizontal line is the mean value of the last $25$
    points. It indicates approximate asymptotic level of
    correlation.}
  \label{Fig003} 
\end{figure}

\begin{figure}
  \resizebox{\hsize}{!}{\includegraphics{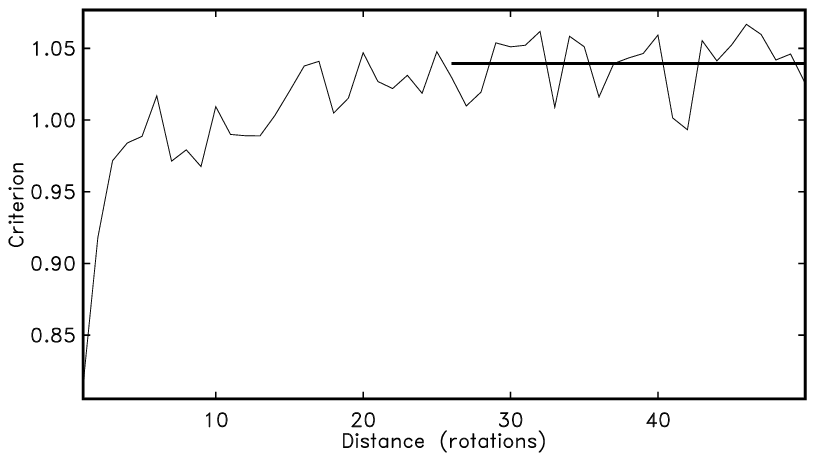}}
  \caption{Statistic $\bar C(l)$ computed for the Northern latitude strip
    $9.9-14.4$.}
  \label{Fig004} 
\end{figure}

\begin{figure}
  \resizebox{\hsize}{!}{\includegraphics{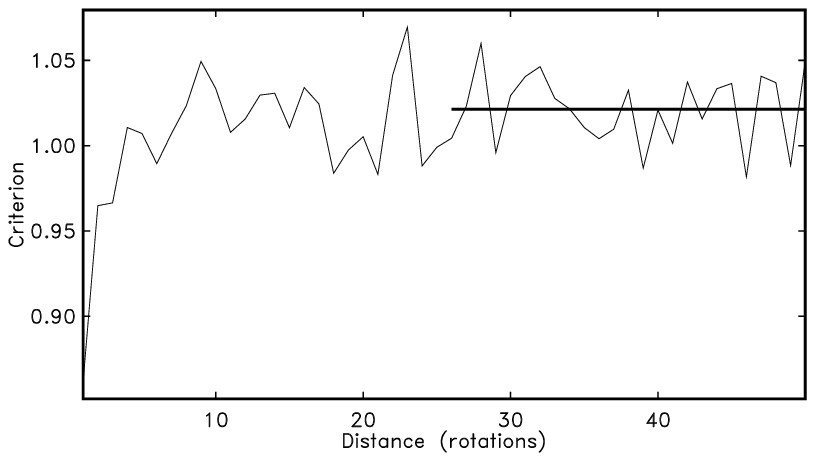}}
  \caption{Statistic $\bar C(l)$ computed for the Northern latitude strip
    $14.5-20.0$.}
  \label{Fig005} 
\end{figure}

\begin{figure}
  \resizebox{\hsize}{!}{\includegraphics{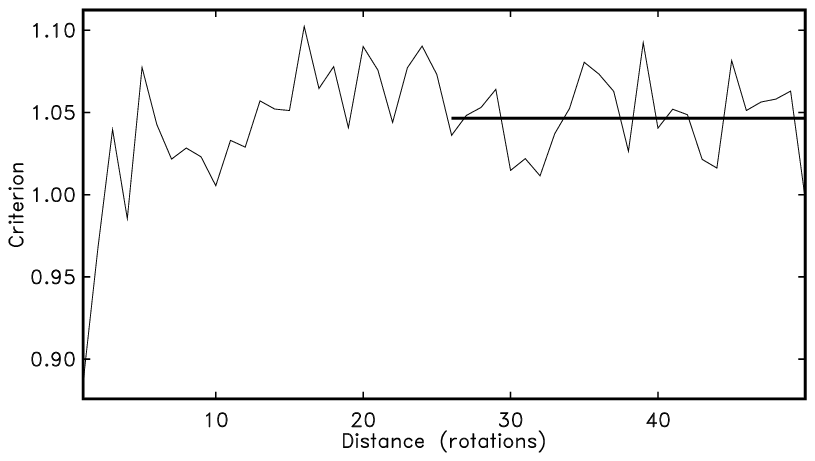}}
  \caption{Statistic $\bar C(l)$ computed for the Northern latitude strip
    $20.1-60.0$.}
  \label{Fig006} 
\end{figure}

\begin{figure}
  \resizebox{\hsize}{!}{\includegraphics{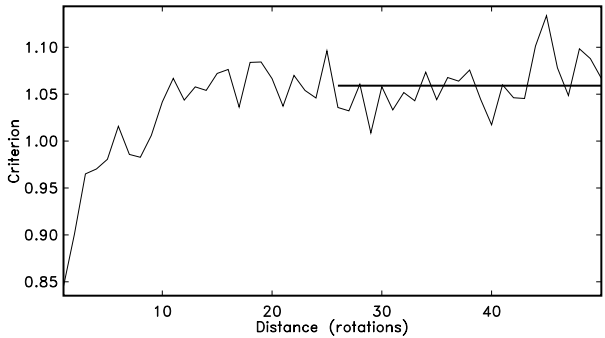}}
  \caption{Statistic $\bar C(l)$ computed for the Southern latitude strip
    $0.0-9.8$.}
  \label{Fig007} 
\end{figure}

\begin{figure}
  \resizebox{\hsize}{!}{\includegraphics{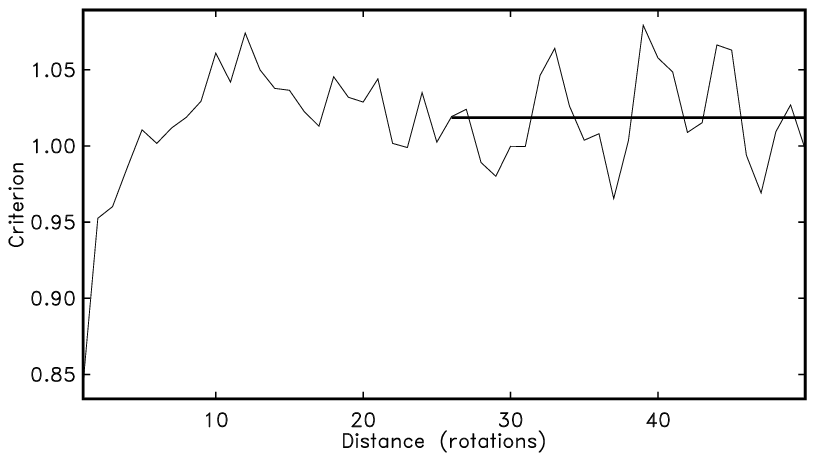}}
  \caption{Statistic $\bar C(l)$ computed for the Southern latitude strip
    $9.9-14.4$.}
  \label{Fig008}
\end{figure}

\begin{figure}
  \resizebox{\hsize}{!}{\includegraphics{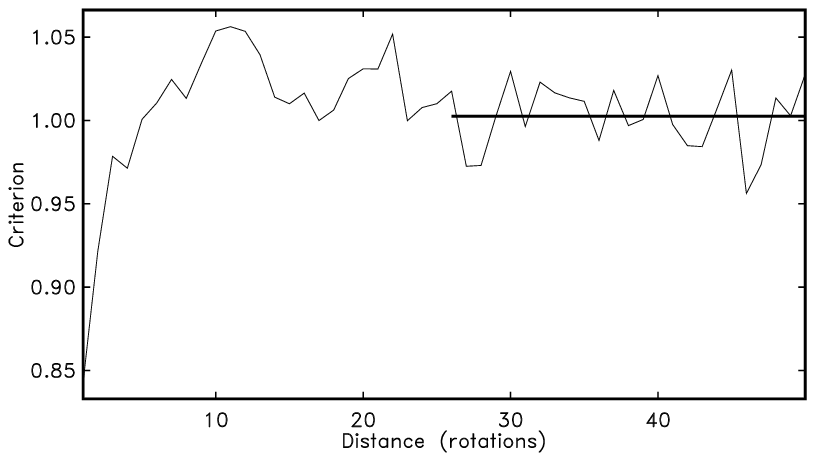}}
  \caption{Statistic $\bar C(l)$ computed for the Southern latitude strip
    $14.5-20.0$.}
  \label{Fig009} 
\end{figure}

\begin{figure}
  \resizebox{\hsize}{!}{\includegraphics{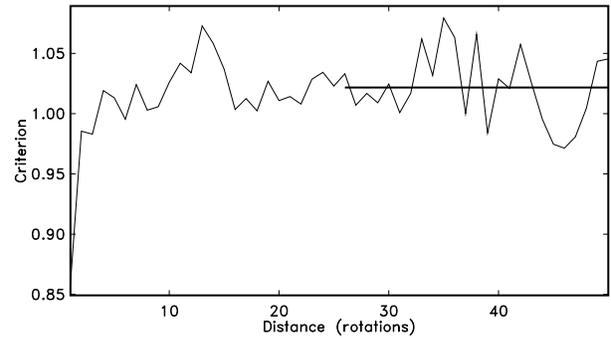}}
  \caption{Statistic $\bar C(l)$ computed for the Southern latitude strip
    $20.1-60.0$.}
  \label{Fig010}
\end{figure}

The results of this kind of analysis are displayed on
Figs.~\ref{Fig003} to \ref{Fig010}. 
On these plots the thick lines indicate mean level over the last 25
points, which gives an asymptotic level of
convergence. As can well be seen from the plots, the inherent scatter
of the $\bar C(l)$ curves is quite large; this is due to the physical
variability of the activity level and the roughness of the
statistic. Therefore it is hard to fix the point where the asymptotic
level is achieved. Some aspects of the curves, however, are quite
indicative. 
First -- the shortest decorrelation time is obtained for the
highest latitudes. 
This is obviously because the width (in degrees) of the high altitude
strips is wider and covers very differently rotating spot
groups. Secondly -- the strips nearest to the equator show the longest
correlations. We can quite safely claim that certain level of
correlation is visible up to 
time span of 15 rotations.

Comparing the estimated decorrelation times for different latitude strips
with the break-down times from Table~\ref{Tab1} we can see that for lower
latitudes the decorrelation times obtained from our analysis are much
shorter than the estimated break-down times. For the highest latitude
strip the decorrelation time is of the same order of magnitude than
the break-down time. Part of this effect could be due to the enhanced
diffusion of the field due to the stretching by the shear in angular
velocity; the latitude dependence (shorter decorrelation times at high
latitudes with the largest relative shear) would support this
interpretation. It seems likely, however, that at least at lower
latitudes, stretching and enhanced turbulent diffusion acting on the
magnetic field due to differential rotation are not the only effects
at play.

\section{Discussion}\label{discussion}
To put our results into general context we will compare them with
a sample of previous analyses.

Large amount of the solar variability research is not based on a full
set of sunspot observations but on some aggregated form of data. Most
typically the daily Wolf sunspot numbers are used. For instance
Bogart~(\cite{Bogart}) analysed these numbers using
autocorrelation functions and power spectra.  The major results were
quite similar to ours - the rotation period around 27 days was
detected and persistence of activity zones was claimed to be of the
order of 10 solar rotations. In principle correlation functions and
power spectra can be considered to be parameter-free statistics. The
aggregated nature of the Wolf numbers, however, does not allow to analyse
latitude dependence of the active clusters.

There is a number of analyses which use longitudinal phase
binning of the surface features.  For instance in a series of papers
Bai~(\cite{Bai87},\cite{Bai88}) used comoving frames (as in our work)
to seek rotation velocities which enhance statistical contrast of
longitudinal distribution of solar flares. The transformed longitudes
for each trial rotation velocity were binned into 12 bins and variance of
obtained distributions was computed. The possibility of differential
rotation was not taken into account. 
To study the persistence of particular active regions were visually
tracked and displayed as ``family trees''. In describing his results
the author proposed a general scheme to characterize hierarchical patterns
of solar activity:
\begin{itemize}
\item single events (sunpots, flares),
\item active regions,
\item activity complexes,
\item active zones.
\end{itemize}
The lifetimes of the activity centres increase with hierarchy - from days
to several years. In a later work Bai~(\cite{Bai03}) used 
Rayleigh-type statistics to analyse transformed longitudes. He computed
standard spectra which are sensitive to unimodal distributions and
spectra which are sensitive to bimodal distributions. As a result he
found that some characteristics of the longitude distributions are
rather persistent in time, even up to decades. Our results
describe average behaviour of the solar activity and consequently some
long-lived elements do not have strong influence, as they are mixed
with other elements whose lifetimes are shorter. It should be also
stressed that $\bar D$ and $\bar C(l)$ statistics do not depend on any
assumption about modality of the underlying variability. All the
``modes'' are automatically accounted for.

Probably the most popular method to study the kinematics of the solar surface
features is a standard power spectrum analysis and its
variants (just an example - Temmer et al.
~\cite{Temmer04}, Giordano~\cite{Giordano}). This kind of analysis can
be applied to latitude strips and in this way the differential
rotation can be taken into account.  From the first sight Fourier
analysis seems to be essentially nonparametric. However, the fact that
it uses single harmonics as base functions prescribes certain form
of preferred activity distributions. The results of the Fourier
methods are often given as a list of certain periods which show up in
power spectra or on wavelet plots. 
The periodicity claim itself is quite a strong statement, as it is often
very hard to find physically solid timing mechanisms for periods
which strongly differ from the obvious one - that of solar rotation.

We want to stress here that in the proposed statistical method no
assumptions about the particular form of the activity indicator
distributions are made. Even more - the statistics $\bar D$ and $\bar
C(l)$ are not seeking certain clusters or other kind of patterns, they
are just used to check whether the ``birth places'' of surface
elements are correlated or not. 
This makes the new method somewhat
similar to the method of ``family trees'' (Bai~\cite{Bai03}) or
longitude-time diagrams (Brouwer \& Zwaan~\cite{Brouwer}).
 
The literature about the longitudinal distribution of solar activity
indicators is so wide that it is not reasonable to compare our results
with all of them. 
It suffices to state that the general patterns revealed so far are
quite similar to those described above.  
The major shortcomings of the previously used methods include the
dependence of the results on some prefixed parameters or on the choice
of a particular distribution model.

\section{Conclusions}\label{conclusions}
When introducing a new method to analyse solar activity patterns 
we started from certain methodological principles:
\begin{itemize}
\item Input data must be homogeneous, comprehensive and cover as long
time base as possible.
\item The analysis method must be free from any prefixed constants.
\item The method must not depend on the model of the activity
  indicator distributions (unimodal, bimodal etc.).
\item The computations must be as simple as possible. 
\end{itemize}
The results obtained using the new method can be ranked using these
underlying principles.  
We start from the most evident and
methodologically ``clean'' facts and proceed towards the statements
which can be doubted or refined using additional devices.
\begin{itemize}
\item The distribution of sunspots is determined by the underlying
  large-scale mechanism which is more persistent than sunspots
  themselves. This shows up as a tendency of new sunspots to occur
  near the places where the previous sunspots were observed.
\item The mean rotation velocity of the large-scale features for the
  Northern hemisphere is $14.35\pm 0.02$ and for the Southern
  hemisphere $14.40\pm 0.02$.
\item The rotation velocities for Northern and Southern hemispheres differ
  slightly. Consequently both velocities manifest certain statistical
  averages and do not suggest a strong meridional coupling.
\item  The large-scale patterns of activity take part in differential
  rotation. The differential rotation curve is somewhat
  shallower if to compare with curves obtained from sunspot tracking (see
  Zapall\`a \& Zuccarello~\cite{Zapalla}).
\item The differential rotation for the Southern hemisphere is more
  similar to that obtained from sunspot analysis.  Differential
  rotation curve for the Northern hemisphere
  deviates from the general rotation law especially near the equator.
\item The strong tendency for the spot groups
to cluster on certain longitude dies off with time. The longest
observable correlations can reach 15 Carrington rotations.
\item The correlations between rotations are more
  pronounced for lower latitudes.
\item The observation of the spot-generating mechanism being affected
  by differential rotation is suggestive of phase mixing occurring in
  the solar convection zone; such a phenomenon is not predicted by
  conventional mean-field dynamo theory.
\end{itemize}
The set of formulated results is impressive if to take into
account the simplicity of the analysis method used.
\begin{acknowledgements}
Part of this work was supported by the Estonian Science Foundation
grant No. 6813 and Academy of Finland grant No. 112020.
\end{acknowledgements}

\end{document}